\documentclass[letterpaper,11pt]{article}
\pdfoutput=1 

\usepackage{graphicx}
\usepackage{natbib}
\usepackage[breaklinks,colorlinks,citecolor=black,linkcolor=black,urlcolor=black]{hyperref}
\usepackage[small,bf]{caption}

\makeatletter
\renewcommand{\section}{\@startsection%
{section}{1}{0mm}{-0.2\baselineskip}%
{0.1\baselineskip}{\normalfont\large\bfseries}}%
\renewcommand{\subsection}{\@startsection%
{subsection}{1}{0mm}{0.01\baselineskip}%
{0.01\baselineskip}{\normalfont\mdseries\bfseries}}%
\makeatother

\usepackage{color}

\newcommand*{\unit}[1]{\ensuremath{\mathrm{\, #1}}}
\newcommand*{\mysub}[2]{\ensuremath{#1_{\mathrm{#2}}}}

\newcommand*{\erg}{\unit{erg}}
\newcommand*{\keV}{\unit{keV}}

\newcommand*{\second}{\unit{s}}
\newcommand*{\cm}{\unit{cm}}

\newcommand*{\E}[1]{\ensuremath{\times 10^{#1}}}

\newcommand*{\ltsim}{\ {\raise-.75ex\hbox{$\buildrel<\over\sim$}}\ }
\newcommand*{\gtsim}{\ {\raise-.75ex\hbox{$\buildrel>\over\sim$}}\ }
\newcommand*{\proptosim}{\ {\raise-.75ex\hbox{$\buildrel\propto\over\sim$}}\ }

\newcommand*{\Omegam}{\mysub{\Omega}{m}}
\newcommand*{\Omegab}{\mysub{\Omega}{b}}

\newcommand*{\Omegal}{\ensuremath{\Omega_{\Lambda}}}

\newcommand*{\LCDM}{\ensuremath{\Lambda}CDM}
\newcommand*{\fgas}{\mysub{f}{gas}\,}
\newcommand*{\Mgas}{\mysub{M}{gas}}

\newcommand*{\dA}{\mysub{d}{A}}

\newcommand*{\omegaDE}{\mysub{\Omega}{DE}}
\newcommand*{\OmegaDE}{\omegaDE}

\usepackage{bm}

\begin{document}
\pagestyle{plain}
\pagenumbering{arabic}

\begin{center} 
\uppercase{\Large \bf{Measuring cosmic distances\\ with galaxy clusters}}
\vspace{0.5cm}
\\
\large S.\,W. Allen$^{1,2,3}$, A.\,B. Mantz$^{4,5}$,  R.\,G. Morris$^{1,3}$, D.\,E. Applegate$^{6}$, \\P.\,L. Kelly$^{7}$, A. von der Linden$^{1,2,8}$, D.\,A. Rapetti$^{8}$, R.\,W. Schmidt$^{9}$
\end{center}

\noindent $^1${Kavli Institute for Particle Astrophysics and Cosmology}

\noindent $^2${Department of Physics, Stanford University, 382 Via Pueblo Mall, Stanford, CA 94305, USA}

\noindent $^3${SLAC National Accelerator Laboratory, 2575 Sand Hill Road, Menlo Park, CA 94025, USA}

\noindent $^4${Kavli Institute for Cosmological Physics}

\noindent $^5${Department of Astronomy and Astrophysics, University of Chicago, 5640 South Ellis Avenue, Chicago, IL 60637, USA}

\noindent $^6${Argelander-Institute for Astronomy, Auf dem H\"ugel 71, D-53121 Bonn, Germany}

\noindent $^7${Department of Astronomy, University of California, Berkeley, CA 94720-3411, USA}

\noindent $^8${Dark Cosmology Centre, Niels Bohr Institute, University of Copenhagen, Juliane Maries Vej 30, DK-2100 Copenhagen, Denmark}

\noindent $^9${Astronomisches Rechen-Institut, Zentrum f\"ur Astronomie der Universit\"at Heidelberg, M\"onchhofstrasse 12-14, 69120 Heidelberg, Germany}

\vspace{0.5cm}
\noindent
\textbf{Abstract:} 

\noindent In addition to cosmological tests based on the mass function
and clustering of galaxy clusters, which probe the growth of cosmic
structure, nature offers two independent ways of using clusters to
measure cosmic distances.  The first uses measurements of the X-ray
emitting gas mass fraction, which is an approximately standard
quantity, independent of mass and redshift, for the most massive
clusters. The second uses combined millimeter (mm) and X-ray
measurements of cluster pressure profiles. We review these methods,
their current status and the prospects for improvements over the next
decade. For the first technique, which currently provides comparable
dark energy constraints to SN\,Ia studies, improvements of a factor of
6 or more should be readily achievable, together with tight
constraints on $\Omegam$ that are largely independent of the
cosmological model assumed. Realizing this potential will require a
coordinated, multiwavelength approach, utilizing new cluster surveys,
X-ray, optical and mm facilities, and a continued emphasis on improved
hydrodynamical simulations.

\vspace{0.5cm}
\section{Introduction}
\vspace{0.1cm}

\noindent Cosmological tests based on observations of galaxy clusters
have seen tremendous improvement in recent years
(\citealt{Allen1103.4829}; Fig.~\ref{fig:contours}a), setting
competitive constraints on cosmological parameters including the
amplitude of the matter power spectrum ($\sigma_8$), the mean matter
and dark energy densities ($\Omega_{\mathrm{m}}$ and
$\Omega_{\mathrm{DE}}$), the dark energy equation of state, $w$
\citep{Allen0706.0033, Vikhlinin0812.2720,
Mantz0909.3098, Rozo0902.3702, Benson1112.5435}, departures from General
Relativity on cosmological scales
\citep{Schmidt0908.2457,Rapetti0911.1787,Rapetti1205.4679}, and the
species-summed neutrino mass \citep{Mantz0911.1788,Reid0910.0008}.

In addition to tests based on measurements of the mass function and
clustering of galaxy clusters \citep{Vikhlinin0812.2720,
Mantz0909.3098, Rozo0902.3702,  Weinberg1201.2434, Benson1112.5435}, which
are discussed elsewhere in these white papers, are two complementary
ways of using galaxy clusters to measure cosmic distances. Both of
these techniques use X-ray observations. The first and most powerful
employs measurements of the X-ray gas mass fraction in clusters, which
is a standard quantity for massive clusters, with minimal scatter from
system to system ({\citealt{Allen0405340,
Allen0706.0033,Battaglia1209.4082,Mantz13}).  The second technique
employs combined X-ray and mm-wavelength measurements of the thermal
pressure profiles in clusters, utilizing the Sunyaev-Zel'dovich effect
\citep{Carlstrom02}.

This white paper briefly describes these two methods to measure
cluster distances, their current status, and the prospects for
improvement over the next decade.

\begin{figure}[t]
  \centering
  \begin{minipage}[c]{0.49\textwidth}
    \includegraphics[width=0.9\textwidth]{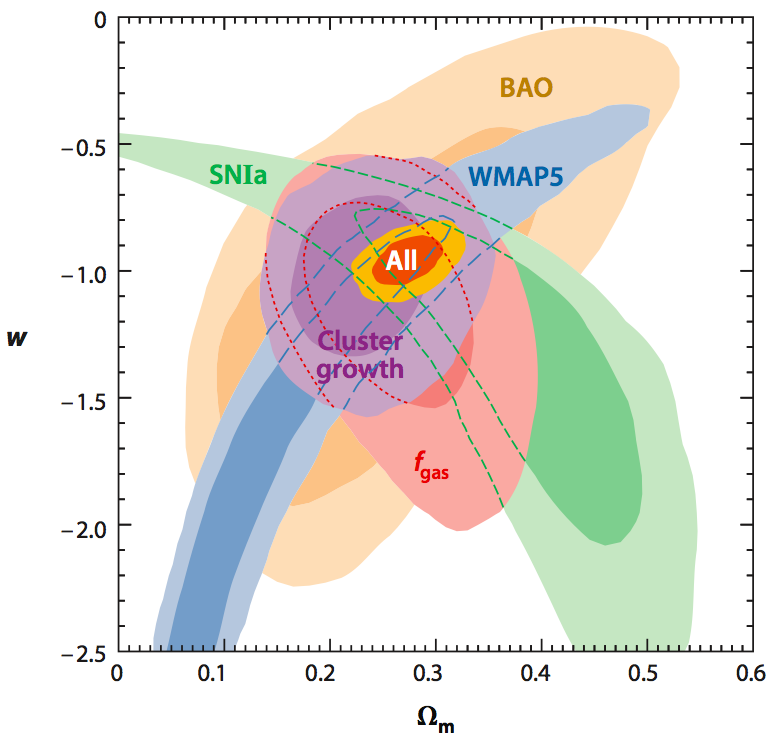}
  \end{minipage}%
  \begin{minipage}[c]{0.49\textwidth}
    \includegraphics[width=0.85\textwidth]{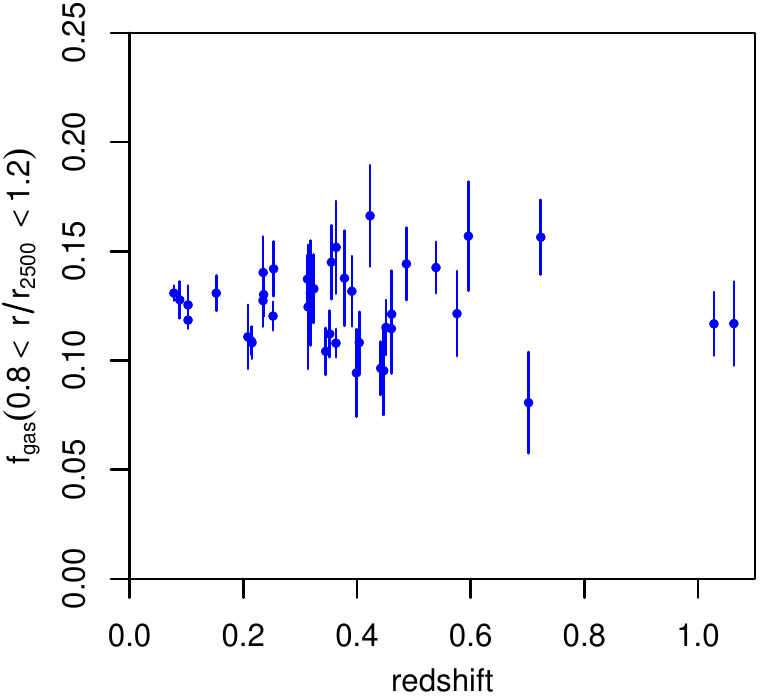}
  \end{minipage}%
 \caption{(\textit{a; left panel}) Cosmological
constraints from galaxy clusters and other probes.  Joint 68.3\% and
95.4\% confidence constraints on the dark energy equation of state,
$w$, and mean matter density, $\Omega_{\mathrm{m}}$, from the observed
abundance and growth galaxy clusters \citep{Mantz0909.3098} and
cluster \fgas~measurements \citep{Allen0706.0033}, compared with those
from the Wilkinson Microwave Anisotropy Probe (WMAP;
\citealt{Dunkley0803.0586}), type Ia supernovae (SN\,Ia;
\citealt{Kowalski0804.4142}) and baryon acoustic oscillations (BAO;
\citealt{Percival0907.1660}) for spatially flat, constant $w$ models.
Gold contours show the results from the combination of all data sets
\citep{Allen1103.4829}. (\textit{b; right panel}) \fgas measurements
in the 0.8--1.2\,$r_{2500}$ shell for the 40 hottest ($kT\gtsim
5$keV), most dynamically relaxed clusters known \citep{Mantz13}. For
illustration, cosmology-dependent quantities are plotted (for an
$\Omegam=0.3$, $\Omegal=0.7$, $h=0.7$ $\Lambda$CDM reference
cosmology), although in practice model predictions are compared with
cosmology-independent measurements.}
\label{fig:contours}
\end{figure}

\vspace{0.5cm}
\section{Distance measurements from the cluster 
X-ray gas mass fraction}
\vspace{0.2cm}

\subsection{Overview and current status}
\vspace{0.2cm}

\noindent The matter content of the most massive galaxy clusters is
expected to provide an almost fair sample of the matter content of the
Universe. The ratio of baryonic-to-total mass in these clusters should,
therefore, closely match the ratio of the cosmological parameters
$\Omega_{\rm b}/\Omega_{\rm m}$. The baryonic mass content of clusters
is dominated by the X-ray emitting gas, the mass of which exceeds the
mass in stars by an order of magnitude (e.g. \citealt {Giodini09})
with other sources of baryonic matter being negligible. The
combination of robust measurements of the baryonic mass fraction in
clusters from X-ray observations, with a determination of $\Omega_{\rm
b}$ from cosmic microwave background (CMB) data or big-bang
nucleosynthesis calculations and a constraint on the Hubble constant,
can therefore be used to measure $\Omega_{\rm m}$
(e.g. \citealt{White93}). This method currently provides one of our
best constraints on $\Omega_{\rm m}$, and is simple and robust in
terms of its underlying assumptions.

Measurements of the apparent evolution of the cluster X-ray gas mass
fraction, hereafter $f_{\rm gas}$, can also be used to probe the
acceleration of the Universe
(\citealt{Allen0405340,Allen0706.0033,LaRoque0604039,Ettori09,Mantz13}).
This constraint originates from the dependence of the $f_{\rm gas}$
measurements, which derive from the observed X-ray gas temperature and
density profiles, on the assumed distances to the clusters, $f_{\rm
gas} \propto d^{1.5}$. The expectation from hydrodynamical simulations
is that for the hottest, most massive systems, and for measurement
radii (corresponding to a given overdensity\footnote{The overdensity,
or ratio of the mean matter density enclosed by a sphere to the
critical density of the Universe, provides a convenient way to define
characteristic radii of clusters. We typically write $M(<r_\Delta) =
(4\pi/3) \, \Delta \, \rho_\mathrm{cr}(z) \, r_\Delta^3$, with, for
example, $r_{2500}$ being roughly $1/4$ the radius of the virialized
region.}) beyond the innermost core ($r \gtsim r_{2500}$), \fgas{}
should be approximately constant with redshift
(\citealt{Battaglia1209.4082, Planelles1209.5058}; and references
therein). Together, the expected behavior of \fgas(z) and distance
dependence of the measurements allow the distances to clusters to be
inferred.

The first detection of cosmic acceleration using the $f_{\rm gas}$
technique was made by \citet{Allen0405340}. This work was extended by
\citet[][see also \citealt{LaRoque0604039,Ettori09}]{Allen0706.0033}
and most recently by \citet{Mantz13} using Chandra X-ray observations
of 40 hot ($kT\gtsim 5$keV), massive, dynamically relaxed clusters
spanning the redshift range $0<z<1.1$. The total Chandra exposure used
by \citet{Mantz13}, after all screening procedures are applied, is
$\sim 3.1$\,Ms. The $f_{\rm gas}$ measurements from the
\citet{Mantz13} study are shown in
Fig.~\ref{fig:contours}b. Fig.~\ref{fig:curres} shows the constraints
from these data for a non-flat \LCDM{} model (left panel; $\Omegam =
0.31 \pm 0.04$, $\Omega_{\Lambda} = 0.69^{+0.17}_{-0.24}$) and a flat,
constant-$w$ model (right panel; $\Omegam = 0.31 \pm 0.04$, $w =
-0.99^{+0.26}_{-0.32}$).  In both cases, the results are marginalized
over conservative systematic uncertainties (Section
\ref{section:priors}). The \fgas{} constraints provide
comparable constraints on dark energy to current SN\,Ia measurements
\citep{Suzuki1105.3470}, and an impressively tight constraint on
$\Omegam$.

\begin{figure}[t] \centering
  \begin{minipage}[c]{0.49\textwidth}
    \includegraphics[width=\textwidth]{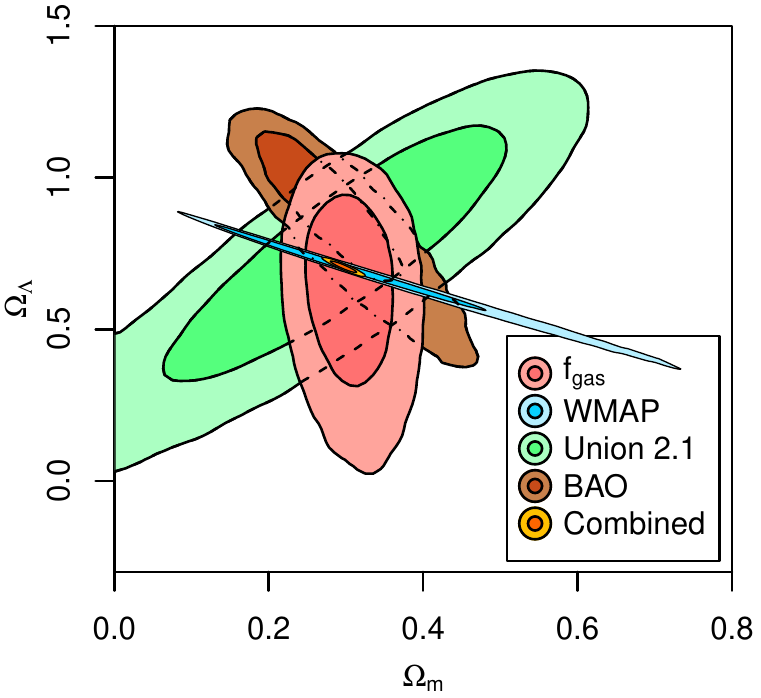}
  \end{minipage}%
  \begin{minipage}[c]{0.49\textwidth}
    \includegraphics[width=\textwidth]{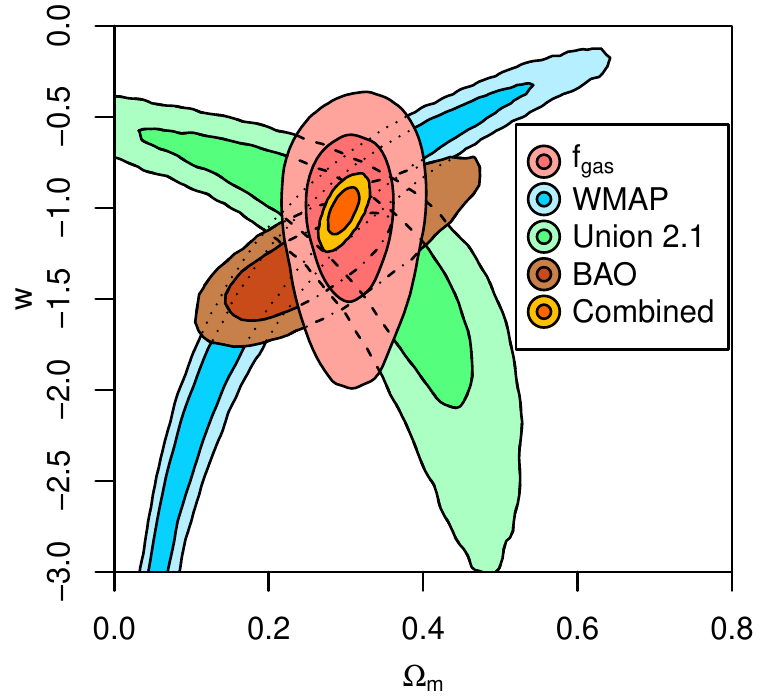}
  \end{minipage}%
  \caption{Joint 68.3\% and 95.4\% confidence constraints for non-flat
\LCDM{} (left) and flat $w$CDM (right) models.  Red shading indicates
confidence regions from current \fgas{} data
(\citealt{Mantz13}; employing standard priors on $\Omegab h^2$
and $h$ as described in Section 5.3.1).  Independent constraints from CMB
measurements (blue; \citealt{Hinshaw1212.5226}), SN\,Ia (green;
\citealt{Suzuki1105.3470}), and BAO (brown; \citealt{Anderson1303.4666}) are also shown.
}
\label{fig:curres}
\end{figure}

\vspace{0.5cm}
\subsection{Modeling the \fgas{} data}
\vspace{0.1cm}

\noindent We define \fgas{} as the ratio of the X-ray emitting gas
mass to the total mass of a cluster, measured at a given overdensity.
This quantity can be determined directly from X-ray observations under
the assumptions of spherical symmetry and hydrostatic equilibrium. To
ensure that these assumptions are as accurate as possible, it is
essential to limit the \fgas{} analysis to the most dynamically
relaxed clusters (\citealt{Allen0405340,Allen0706.0033,Mantz13}).  A
further restriction to the hottest, most X-ray luminous
systems simplifies the cosmology analysis and minimizes the required
exposure times (see also Section~\ref{section:optimal}).

\citet{Mantz13} use Chandra \fgas{} measurements for a sample of 40 of
the hottest ($kT_{2500}> 5$\,keV), most dynamically relaxed clusters
known. The selection on dynamical relaxation state is based on
quantitative morphological criteria computed from X-ray images
(Section~\ref{section:optimal}).  The \fgas{} measurements are made
within a spherical shell spanning the radial range
0.8--1.2\,$r_{2500}$ (angular range 0.8--1.2\,$\theta^{\rm
ref}_{2500}$), determined for a chosen reference cosmology.

Hydrodynamical simulations, tuned to match the observed thermodynamic
properties and scaling relations of galaxy clusters
(e.g. \citealt{Battaglia1209.4082, Planelles1209.5058}) suggest that
for the hottest, most massive clusters, and for measurement radii
beyond the innermost core, the true gas mass fraction at a given
overdensity should be approximately constant in mass and redshift,
\begin{equation} f^{\rm }_{\rm gas}(z;\theta^{\rm
}_{2500})=\Upsilon_{2500}\,\left(\frac{\Omega_{\rm b}}{\Omega_{\rm
m}}\right),
\end{equation} where $\Upsilon_{2500}$, also known as the gas
depletion parameter, is the average ratio of the cluster gas mass
fraction to the cosmic mean baryon fraction at the appropriate radii.

In comparing with data, an angular correction factor, $A$, is required
to account for the fact that $\theta^{\rm ref}_{2500}$ for the
reference cosmology and $\theta^{\rm }_{2500}$ for a given trial
cosmology are not identical. Observations show that over the radial
range of interest, the $f_{\rm gas}(r)$ profiles of hot, relaxed
clusters can be described by a power-law model, independent of the
reference cosmology \citep{Allen0706.0033}.  The \fgas{} measurements
in the reference cosmology can then be related to a predicted curve
for a particular trial cosmology, $f^{\rm ref}_{\rm gas}(z)$, through
the relation

\begin{equation}
  f^{\rm ref}_{\rm gas}(z;\theta^{\rm ref}_{2500}) = 
A\, \Upsilon_{2500}\,\left(\frac{\Omega_{\rm b}}{\Omega_{\rm m}}\right)
\left(\frac{d^{\rm ref}_{\rm A}}{d_{\rm A}}\right)^{3/2},
\label{eq:fgas}
\end{equation}
where
\begin{equation}
A =  \left(\frac{\theta^{\rm ref}_{2500}}{\theta^{\rm }_{2500}}\right)^{\eta} \sim \left(\frac{[H(z)\,d_{\rm A}(z)]^{\rm }}{[H(z)\,d_{\rm A}(z)]^{\rm ref}}\right)^{\eta}\,.
\label{eq:angcosm}
\end{equation}

\noindent For the 0.8--1.2\,$\theta_{2500}$ shell, $\eta=0.442 \pm
0.035$ \citep{Mantz13}. The explicit appearance of \Omegam{} in
equation~(\ref{eq:fgas}) means that low-redshift clusters (for which
the dependence of $\dA$ on the details of dark energy is negligible),
in combination with external priors on $\Omegab h^2$ and $h$, provide
a tight and essentially model-independent constrain on \Omegam{}.

Following \citet{Allen0706.0033}, we extend equation (\ref{eq:fgas})
to also include allowances for systematic uncertainties,

\begin{equation}
  f^{\rm ref}_{\rm gas}(z;\theta^{\rm ref}_{2500}) = 
K\, A\, \Upsilon_{2500}\,\left(\frac{\Omega_{\rm b}}{\Omega_{\rm m}}\right)
\left(\frac{d^{\rm ref}_{\rm A}}{d_{\rm A}}\right)^{3/2}\,.
\label{eq:fgas3}
\end{equation}

\noindent Here $K$ is a systematic uncertainty on the overall
normalization of the curve, encompassing terms relating to (among
other factors) instrumental calibration and the expected mean level of
non-thermal pressure support in the measurement shell \citep{Mantz13}.
Fortunately, the value of $K$ can be constrained robustly through
combination with independent weak lensing mass measurements for the
target clusters, with current uncertainties at the ten
per cent level ($K = 0.94 \pm 0.09$; \citealt{Applegate13b}).  Note
that only the mean level of these systematic effects is modeled by
$K$; cluster-to-cluster scatter is manifested as intrinsic scatter in
the measurements.

The gas depletion parameter, $\Upsilon$, reflects the thermodynamic
history of the X-ray emitting cluster gas over the history of cluster
formation. Systematic uncertainties are present in both the
normalization and evolution of this parameter, although these
uncertainties are minimized by our use of a measurement shell as
opposed to traditional measurements within enclosed spheres
(i.e. including cluster centers).\footnote{Systematic uncertainties in
the prediction of $\Upsilon$, which are primarily related to
uncertainties in the baryonic physics, mainly affect the innermost
regions of clusters at $r<0.5r_{2500}$.} Defining a simple
evolutionary model, $\Upsilon_{2500}=\Upsilon_{\rm 0}(1+\alpha_{\rm
\Upsilon}z)$, recent hydrodynamic simulations spanning a range of
plausible astrophysical models \citep{Battaglia1209.4082,
Planelles1209.5058} show that, for the hottest clusters and
measurement shells 0.8--1.2\,$r_{\rm 2500}$, the full range of
uncertainty is conservatively spanned by $\Upsilon_0 = 0.845 \pm
0.042$ and $\alpha_{\rm \Upsilon} = 0.00\pm0.05$ (uniform priors).

We note that it is straightforward to additionally model the intrinsic
scatter in \fgas{} measurements as well as a possible dependence of
the mean $\fgas(z)$ on mass, constraining these terms simultaneously
with the cosmological parameters. \citet{Mantz13} place constraints on
the intrinsic scatter (see Section 5.1), and confirm the expectation
that the mean \fgas{} in the 0.8--1.2\,$r_{2500}$ shell is independent
of mass.

\vspace{0.5cm}
\section{Distance measurements from SZ and X-ray pressure profiles}
\vspace{0.2cm}
\subsection{Overview and current status}
\vspace{0.2cm}

\noindent Cosmic microwave background (CMB) photons passing through a
galaxy cluster have a non-negligible chance to inverse Compton scatter
off the hot, X-ray emitting gas.  This scattering boosts the photon
energy and gives rise to a small but significant frequency-dependent
shift in the CMB spectrum observed through the cluster, known as the
thermal Sunyaev-Zel'dovich (hereafter SZ) effect \citep{Sunyaev72}.

\citet{WhiteSilk78} noted that X-ray and SZ measurements could in
principle be combined to determine distances to galaxy clusters. The
spectral shift to the CMB due to the SZ effect can be written in terms
of the Compton $y$-parameter, which is a measure of the integrated electron
pressure along the line of sight, $y \propto \int n_{\rm e}\, T\, dl$
. Given an observed SZ signal at mm or radio wavelengths, and a
predicted SZ signal from X-ray measurements of the gas density and
temperature profiles then, given the distance dependence of the X-ray
measurements, we can solve for $\dA$.

This test, sometimes referred to as the XSZ (or SZX) test, is
intrinsically less powerful than the \fgas{} test due to the
cosmological signal being proportional to $\dA^{\,0.5}$, rather than
$\dA^{\,1.5}$.\footnote{As discussed in this document, the
\fgas{} test also provides a second, independent constraint on
$\Omega_{\rm m}$ from the fair sample argument.}  As a result, the XSZ
test has to date only been used to constrain the Hubble
parameter. \citet{Bonamente0512349} used this method to measure the
distances to 38 X-ray luminous clusters spanning the redshift range
$0.14 < z < 0.89$.  Assuming spatial flatness and fixing $\Omega_{\rm
m} = 0.3$, they determined a Hubble parameter, $h =
0.77^{+0.11}_{-0.09}$, consistent with the results from the 
Hubble Key Project ($h = 0.72\pm0.08$;
\citealt{Freedman0012376}).

\vspace{0.5cm}
\subsection{Modeling the XSZ data}
\vspace{0.1cm}

\noindent For the true, underlying cosmology, the measurements of the
Compton $y$-parameter from the X-ray and mm-wavelength data should be
equal.  For a given cosmology, the $y$-parameter determined from the
X-ray data depends on the square root of the angular diameter distance
to the cluster, whereas the observed SZ flux at mm (or radio)
wavelengths is independent of the cosmology assumed. Combining these
$y$-parameter results, we can measure the distances to the clusters as
a function of redshift:
 
\begin{equation}
{y_{\rm X}^{\rm ref}} = {y_{\rm SZ}^{\rm }}\, k(z)\left( \frac{d_{\rm A}^{\rm ref}}{d_{\rm A}^{\rm }}\right)^{1/2}\,.
\label{eq:ycompton}
\end{equation}

\noindent Here $y_{X}^{\rm ref}$ is the X-ray measurement of the
$y$-parameter for a reference cosmology and $y_{\rm SZ}$ is the
cosmology-independent mm (or radio) observation. Following a similar
approach to the $f_{\rm gas}$ case, we again incorporate systematic
allowances into equation (\ref{eq:ycompton}), via the term
$k(z)=k_{0}(1+\alpha_{k} z)$, which in this case accounts for the
combined uncertainties due to instrument calibration, geometric
effects, gas clumping and their evolution. Again, only the mean level
of these systematic effects is modeled by $k(z)$; cluster-to-cluster scatter
will be manifested as intrinsic scatter in the measurements.

\vspace{0.5cm}
\section{The optimal cluster targets}
\vspace{0.2cm}
\label{section:optimal}

\noindent A simplifying aspect of the \fgas{} and XSZ methods is that
the optimal targets for both experiments are identical. In each case,
the ideal targets are the most massive -- and therefore hottest and
most X-ray luminous -- dynamically relaxed clusters.

X-ray images with modern X-ray telescopes (point spread functions of
$\sim 10$ arcsec or better) allow for the straightforward assessment
of the dynamical states of clusters based on measurements of the
peakiness and symmetry of the X-ray emission, and the level of
isophote centroid variation. This selection can be easily automated
(e.g. \citealt{Mantz13}).\footnote{In principle, future X-ray missions
with larger collecting area and higher spectral resolution could also
utilize measurements of gas motions within clusters and/or
thermodynamic maps to assess their dynamical states.}  Existing
Chandra and XMM-Newton data for clusters at $0<z<1.1$ indicate that
approximately $1/8$ clusters with $kT>5$ keV can be classified as
highly relaxed \citep{Mantz13}.

From the \fgas{} perspective, the restriction to the most dynamically
relaxed clusters, for which the assumptions of hydrostatic equilibrium
and spherical symmetry should be most valid, minimizes the systematic
scatter in the \fgas{} measurements, which in turn impacts on the
cosmological constraining power. The restriction to the most massive
clusters additionally simplifies the prediction of the gas depletion
factor, $\Upsilon(z)$: for clusters with $kT> 5$keV and for the
measurement radii of interest (0.8--1.2\,$r_{2500}$) the depletion
factor is predicted to be essentially independent of mass and redshift
and to exhibit minimal intrinsic scatter from cluster to cluster
\citep{Battaglia1209.4082, Planelles1209.5058, Mantz13}.
Since the most massive clusters are also the most luminous at a given
redshift, these targets will require the shortest exposure to reach a
given X-ray measurement precision. Likewise, from the XSZ perspective,
the restriction to the hottest, most dynamically relaxed clusters
maximizes the X-ray and mm-wavelength signals and minimizes the systematic
uncertainties associated with geometry and substructure/clumping.

\vspace{0.5cm}
\section{Prospects for improvement}
\vspace{0.2cm}
\subsection{An observing program for cluster distance measurements}
\vspace{0.2cm}

\noindent Future \fgas{} and XSZ experiments will build upon new
cluster surveys.  Together, these surveys will map the whole sky,
finding in excess of 100,000 clusters, including thousands of hot,
massive systems out to high redshifts.  Fig.~\ref{fig:zhist}a shows
the predicted distribution of clusters with temperature $kT>5$ keV as
a function of redshift.

\begin{figure}[t] \centering
  \begin{minipage}[c]{0.49\textwidth}
    \includegraphics[width=0.9\textwidth]{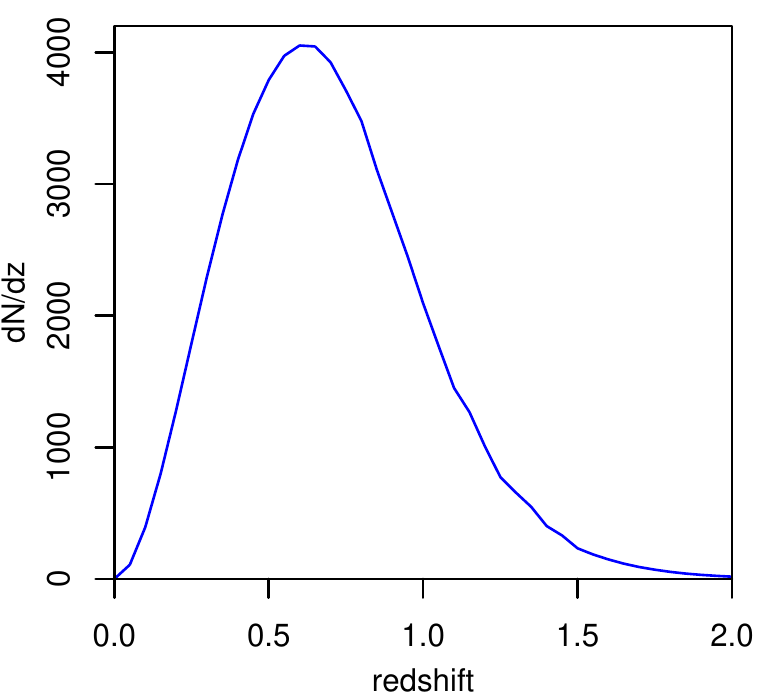}
  \end{minipage}%
  \begin{minipage}[c]{0.49\textwidth}
    \includegraphics[width=0.9\textwidth]{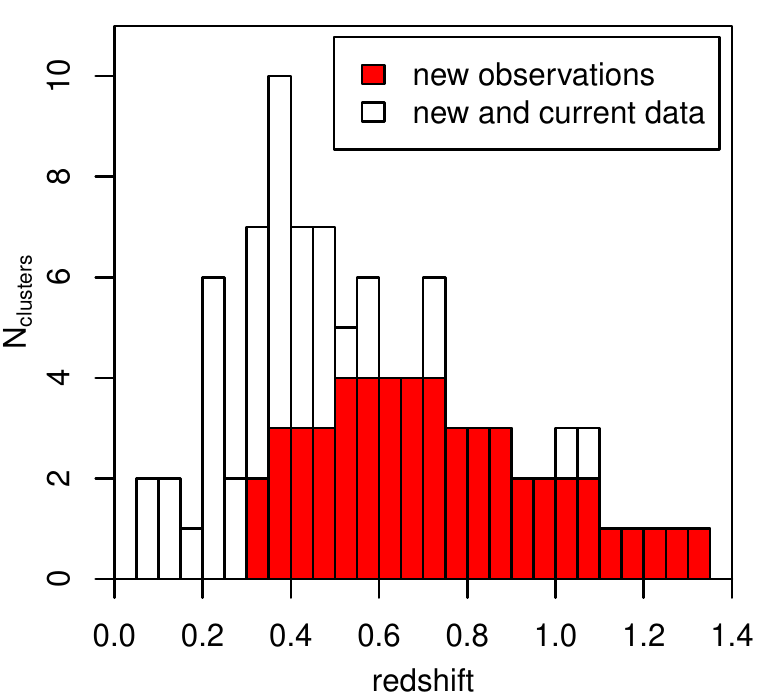}
  \end{minipage}%
  \caption{(\textit{a; left panel}) Differential number of clusters
with temperatures $kT>5\keV$ predicted to be detected by the eROSITA
survey. We assume a concordance \LCDM{} model with the halo mass
function of \citet{Tinker0803.2706}, cluster luminosity and
temperature scaling relations measured by \citet{Mantz0909.3099}, and
a survey flux limit of $3.3\E{-14}\erg\second^{-1}\cm^{-2}$ in the
0.5--2.0\,keV band.  (\textit{b; right panel}) The red histogram shows
a plausible redshift distribution for new cluster observations
achievable with 10\,Ms of Chandra time over the next decade. These
represent a fair sample of the distribution shown in the left panel at
$z\geq0.3$. The union of these future observations with the current
data set of \citet{Mantz13} is shown in the white histogram. This has
been used in projecting future cosmological constraints (Section
5.4).}
\label{fig:zhist}
\end{figure}

After completing its 4-year X-ray survey, the eROSITA mission
\citep{Merloni1209.3114} will spend at least two more years carrying
out targeted, follow-up observations to measure low-scatter X-ray mass
proxies: X-ray temperatures ($kT$), gas masses (\Mgas), and $Y_{\rm
X}$ values (the product of gas mass and temperature;
\citealt{Allen1103.4829}). The addition of such low-scatter mass proxy
measurements for even a small fraction of the clusters in a survey can
in principle boost the cosmological constraining power by a factor of
a few with respect to self-calibration of the mass function and
clustering data alone \citep{Wu10}. Of order a thousand follow-up
measurements are likely to be made with eROSITA, gathering 1000 or
more counts per target. In many cases these observations will also be
sufficient to assess the dynamical states of the clusters.

The hottest, most X-ray luminous, dynamically relaxed clusters
identified in this way will be the targets for further, deeper
follow-up by flagship X-ray observatories and SZ telescopes, designed
to enable the \fgas{} and XSZ experiments.  For simplicity, we assume
here that this deeper X-ray follow-up will be carried out exclusively
with Chandra, utilizing an observing budget of 10\,Ms over the next
decade (approximately 5\% of the available satellite observing
time). 

\citet{Mantz13} determine an intrinsic scatter in \fgas{} measurements
for the hottest, most relaxed clusters at $0<z<1.1$ in the
0.8--1.2\,$r_{2500}$ shell of $7.4\% \pm 2.3$\%. This implies an
intrinsic scatter in the distance measurements to individual clusters
of $\sim 5$\%. Since it is inefficient for \fgas{} measurements of
individual clusters to exceed this precision significantly, we propose
a target precision for future Chandra \fgas{} measurements (in the
same shell) of $\sim 15\%$. The required exposure times for individual
targets can be estimated straightforwardly from existing
data.\footnote{Based on the existing Chandra data, the exposure time, $t$,
needed to reach a given fractional uncertainty on \fgas{},
$\varepsilon$, in the $0.8-1.2r_{2500}$ measurement shell for a
cluster at redshift $z$ is well approximated by $t = \left(
5.65\E{-7}\,\mathrm{ks}\right)\left[ d_\mathrm{L}(z) / \mathrm{Mpc}
\right]^2 \varepsilon^{-2} E(z)^{-2.3}$, where $E(z)=H(z)/H_0$ and
$d_\mathrm{L}(z)$ is the luminosity distance.  }

Fig.~\ref{fig:zhist}b shows the redshift histogram of a possible
target list of hot ($kT>5$ keV), relaxed clusters that could be
observed in 10\,Ms of Chandra time.  The targets are drawn fairly from
the full distribution shown in Fig.~\ref{fig:zhist}a, apart from the
requirement that $z\geq0.3$. The fiducial parameter values used to
simulate these data are shown in Table~\ref{tab:fiducial}.  Exposure
times are tuned to provide a precision in \fgas{} of $\sim 15$\% in
the 0.8--1.2\,$r_{r2500}$ shell, with the maximum exposure per cluster
limited to 300\,ks\footnote{Exposure times of this length are also
well suited to a broad range of astrophysical studies, for example
studies of the incidence, properties and evolution of cool cores,
which may have high priority in the community
(e.g. \citealt{Siemiginowska10,Santos1111.3642,McDonald1305.2915}).
Alternatively, the same 10\,Ms exposure time could be divided among,
e.g., twice as many new clusters, leading to similar cosmological
constraints.}. In total $\sim 53$ new clusters can be observed in this
way.  When combined with current data \citep{Mantz13}, this gives a
sample of $\sim 93$ measurements meeting the target precision over the
redshift range $0<z<1.4$. Our simulations of projected cosmology
constraints for the \fgas{} experiment (Section 5.4) are carried out
for this combined data set.

For the XSZ test, which does not require information on the gas
depletion factor, it is efficient to use a wider shell, encompassing
most of the sphere enclosed within $1.2r_{2500}$ (excluding just a
small central region of radius $r<0.2r_{2500}$). The Chandra X-ray
observations described above, which are intended to measure \fgas{} in
the 0.8--1.2\,$r_{2500}$ shell to $\sim 15\%$ precision, should also
enable $y_{\rm X}^{\rm ref}$ to be measured to $\sim 7.5$\%.  However,
given the larger level of intrinsic scatter expected in the XSZ
measurements ($\sim20\%$, corresponding to 40\% in distance), due
primarily to the impact of triaxiality and projection effects on the
measured $y_{\rm SZ}$ values
\citep{Bonamente0512349,Hallman07,Shaw08}, it is likely that the
shallower eROSITA X-ray data set described above, which will contain
many more clusters, will be at least as useful for this
work. Envisaging some modest restriction in the XSZ target selection
on the basis of dynamical state, we have simulated an XSZ data set
appropriate for 500 eROSITA clusters, fairly sampling the $dN/dz$
curve in Fig.~\ref{fig:zhist}a at redshifts $z>0.1$, with X-ray
measurements made to 20\% precision (roughly the minimum precision
necessary to measure a useful mass proxy). We assume that the
accompanying, targeted mm observations, made with facilities such as 
CARMA\footnote{http://www.mmarray.org/} or
CCAT\footnote{http://www.ccatobservatory.org/}, will have
significantly greater precision. Our projected cosmology constraints
for the XSZ experiment (Section 5.4) are for this data set.

\vspace{0.5cm}
\subsection{Figure of merit}
\vspace{0.2cm}

\noindent We quantify the improvements in cosmological constraining
power with the proposed observing strategy for three cosmological
models: non-flat $\Lambda$CDM, flat $w$CDM (i.e. a flat dark energy
model with a constant equation of state, $w$) and a flat, evolving
dark energy model with $w(a)=w_{\rm 0}+w_{\rm a}(1-a)$, where $a$ is
the scale factor, similar to that employed by the Dark Energy Task
Force (DETF; \citealt{Albrecht0609591}). For each model, our figure of
merit is defined as the inverse of the area enclosed by the 95.4\%
confidence contour for the associated pair of parameters
[($\Omega_{\rm m},\Omega_{\rm \Lambda}$), ($\Omega_{\rm DE},w$) or
($w_0,w_a$), respectively], normalized by the constraints provided by
current data \citep{Mantz13}. We also examine the
fractional uncertainty in the marginalized constraint on \Omegam,
which is well constrained by the data.

As with the real data, our determination of projected constraints
employs the Metropolis Markov Chain Monte Carlo (MCMC) algorithm
implemented in the COSMOMC\footnote{http://cosmologist.info/cosmomc/}
code \citep{Lewis0205436} to determine posterior parameter
distributions.  This ensures maximal consistency in our calculation of
figures of merit, and that we capture fully the various degeneracies
between model parameters and priors (\citealt{Wolz1205.3984}).  Note
that, as a result, our figure of merit differs slightly from that
introduced by the DETF, in which a Gaussian approximation to the
posterior is used to estimate the area enclosed by the confidence
contours. We measure the area enclosed directly from Monte Carlo
simulations of the posterior, including all details of its
shape.\footnote{A previous study of the potential of the \fgas{} and
XSZ experiments, carried out in the context of the planned
Constellation-X and XEUS satellites, was presented by
\citet{Rapetti0710.0440}. Their analysis employed the standard DETF
figure of merit, calculated from MCMC chains, and Planck simulated
data. However, it underestimated the progress in mitigating systematic
uncertainties; in particular, the pessimistic scenarios considered by
those authors can now be excluded.}

\vspace{0.5cm}
\subsection{Priors and systematic allowances}
\vspace{0.2cm}
\label{section:priors}

\begin{table*}
\begin{center}
\caption{Systematic allowances on parameters, expressed as fractions of their fiducial values (see Table~\ref{tab:fiducial}).}
\label{table:sys}
\begin{tabular}{ c c c c c c c }
&&&&&  \\   
Parameter            & & Current             & \multicolumn{2}{c}{Future}   &  Form    \\
& & & pessimistic & optimistic & \\
\hline                                              
\noalign{\vskip 5pt}
\noalign{\vskip 5pt}
                                                 
$\Omega_{\rm b}h^2$ & &  $\pm0.04$  & $\pm0.04$&$\pm0.01$   & Gaussian \\
$h$                 & &  $\pm0.03$  & $\pm0.03$&$\pm0.01$   & Gaussian \\
\noalign{\vskip 10pt}  
$K$                 & &  $\pm0.10$  & $\pm0.05$&$\pm0.02$   & Gaussian \\
$\Upsilon_0$        & &  $\pm0.05$  & $\pm0.05$&$\pm0.02$   & Uniform  \\
$\alpha_{\Upsilon}$ & &  $\pm0.05$  & $\pm0.05$&$\pm0.02$   & Uniform  \\
\noalign{\vskip 10pt}                                                            
$k_0$               & &  $\pm0.12$  & $\pm0.05$&$\pm0.02$   & Gaussian \\
$\alpha_k$       & &    & $\pm0.05$&$\pm0.02$   & Uniform \\
\noalign{\vskip 10pt}                                    
\hline                      
\end{tabular}
\end{center}
\end{table*}

\begin{table*}
\begin{center}
\caption{Fiducial parameter values used to simulate future cluster data.}
\label{tab:fiducial}
\vspace{0.2cm}
\begin{tabular}{c@{ $=$ }c@{\hspace{2em}}c@{ $=$ }c}
  \hline
  $h $&$ 0.7$ & $\Upsilon_0 $&$ 0.845$ \\
  $\Omegab $&$ 0.045$ & $\alpha_\Upsilon $&$ 0.0$ \\
  $\Omegam $&$ 0.3$ & $K $&$ 1.0$ \\
  $\Omegal $&$ 0.7$ & $k_0 $&$ 1.0$ \\
  $w_0 $&$ -1.0$ & $\alpha_k $&$ 0.0$ \\
  $w_a $&$ 0.0$ \\
 \hline
\end{tabular}
\end{center}
\end{table*}

\noindent In order to keep the interpretation of projected results as
simple as possible, we present results for two sets of priors and
systematic allowances, corresponding to pessimistic and optimistic
(standard) scenarios.

\subsubsection{\fgas{} method}

Cosmological analyses based on \fgas{} data alone require external
priors on $\Omega_{\rm b}h^2$ and $h$. The current analysis
\citep{Mantz13} uses $\Omega_{\rm b}h^2 = 0.0223 \pm 0.0009$
\citep{Pettini1205.3785} and $h = 0.738 \pm 0.024$
\citep{Riess1103.2976}. Our pessimistic assumption is that this
will not improve. Optimistically, we assume that CMB and other data
will improve these priors to the $\sim 1\%$ level over the next
decade. Note that when combining \fgas{} and CMB data directly,
external priors on $\Omega_{\rm b}h^2$ and $h$ are not required.

The calibration prior for the \fgas{} analysis, $K$, can be
constrained robustly using weak lensing measurements for the target
clusters.  Current uncertainties in the value of
this prior are at the $10$\% level ($K = 0.94 \pm 0.09$; 
\citealt{Applegate13b}), with no evidence for evolution,
and are limited by the small number of hot, relaxed clusters with
rigorous weak lensing mass measurements.  The path toward improving
this constraint appears straightforward, however, with programs to
gather additional lensing data, and to improve the accuracy of the
shear calibration and lens modeling, underway
\citep{vonderlinden1208.0597,Kelly1208.0602,Applegate1208.0605}.
Optimistically, calibration of $K$ at the 2\% level can be envisaged,
with calibration to at least $5\%$ over the next decade appearing
straightforward.

Systematic uncertainties are present in both the normalization and
evolution of the gas depletion parameter.  Writing
$\Upsilon(z)=\Upsilon_{\rm 0}(1+\alpha_{\rm \Upsilon}z)$, the
hydrodynamic simulations of \citet{Planelles1209.5058} and
\citet{Battaglia1209.4082} show that for the most massive clusters and
for measurement shells 0.8--1.2\,$r_{\rm 2500}$, $\Upsilon_0 = 0.845
\pm 0.042$ and $\alpha_{\rm \Upsilon}=0.00\pm0.05$ (uniform priors;
\citealt{Mantz13}). These priors span the full range of current
systematic uncertainty. Our pessimistic assumption for future analyses
employs these same conservative priors, whereas our optimistic
scenario assumes a 2.5 times reduction in their ranges. We
note that our prior on $\alpha_{\rm \Upsilon}$ also serves as a
catch-all for other small-scale evolutionary biases that may affect
the measured \fgas{} value.

Note that the priors on $K$ and $\Upsilon(z)$ only model systematics
affecting the mean value of $\fgas(z)$; systematics acting
stochastically on a cluster-to-cluster basis will be manifested as
intrinsic scatter in the measurements. Our projections for future
Chandra data assume an intrinsic scatter in \fgas{} of 7.5\% for the
0.8--1.2\,$r_{\rm 2500}$ measurement shell based on current data
($7.4\%\pm2.3\%$; \citealt{Mantz13}), corresponding to an intrinsic
scatter in distance of $\sim 5$\%, in both the pessimistic and
optimistic scenarios.

\subsubsection{XSZ method}

The calibration prior for the XSZ analysis, $k(z)$, is set by the
accuracy of the temperature and flux calibration for the X-ray and mm
instruments. On the X-ray side, the temperature calibration is likely
to see near-term improvements through the availability of independent
temperature measurements based on X-ray emission line ratios, provided
by the ASTRO-H satellite (\citealt{Takahashi1010.4972}; these will
augment the standard temperature estimates based on the shape of the
bremsstrahlung continuum). Gravitational lensing data will also play a
role. Our pessimistic assumption for this calibration prior assumes a
combined systematic uncertainty in future XSZ measurements of 5\%,
shrinking to 2\% in the optimistic case. For the relevant radial range
of relaxed clusters, we do not expect these systematics to be
redshift-dependent, to first order. To be conservative, however, we
also adopt 5\% and 2\% uncertainties on $\alpha_k$ in the pessimistic
and optimistic cases, respectively.  In all cases, we assume an
intrinsic scatter in the XSZ measurements of $20\%$ (corresponding to
40\% in distance; \citealt{Bonamente0512349,Hallman07,Shaw08}).  We
note, however, that this scatter has yet to be investigated in detail
for hot, relaxed clusters specifically, and it is possible that our
assumption may overestimate the true level of intrinsic scatter.

\vspace{0.5cm}
\subsection{Simulation results}
\vspace{0.2cm}

\begin{figure}[t] \centering
  \begin{minipage}[c]{0.33\textwidth}
    \includegraphics[width=\textwidth]{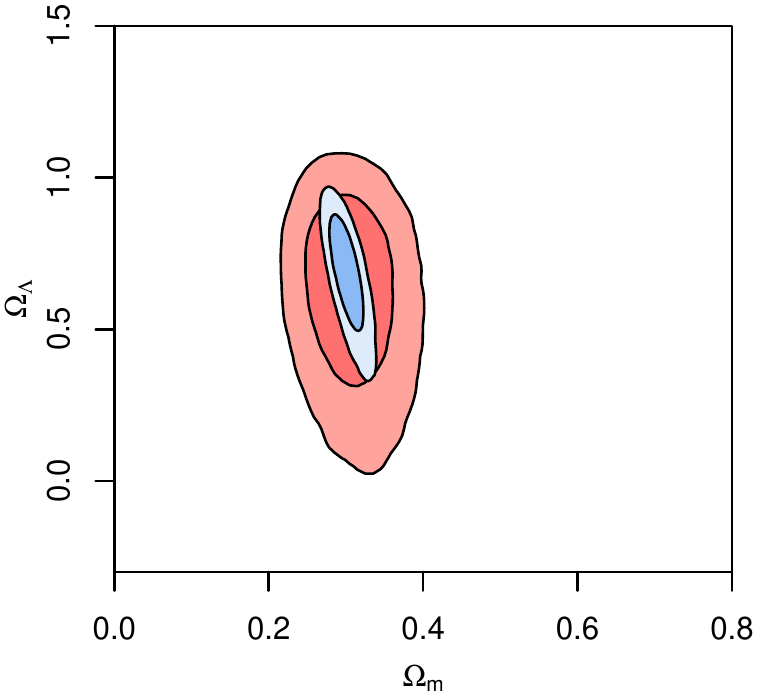}
  \end{minipage}%
  \begin{minipage}[c]{0.33\textwidth}
    \includegraphics[width=\textwidth]{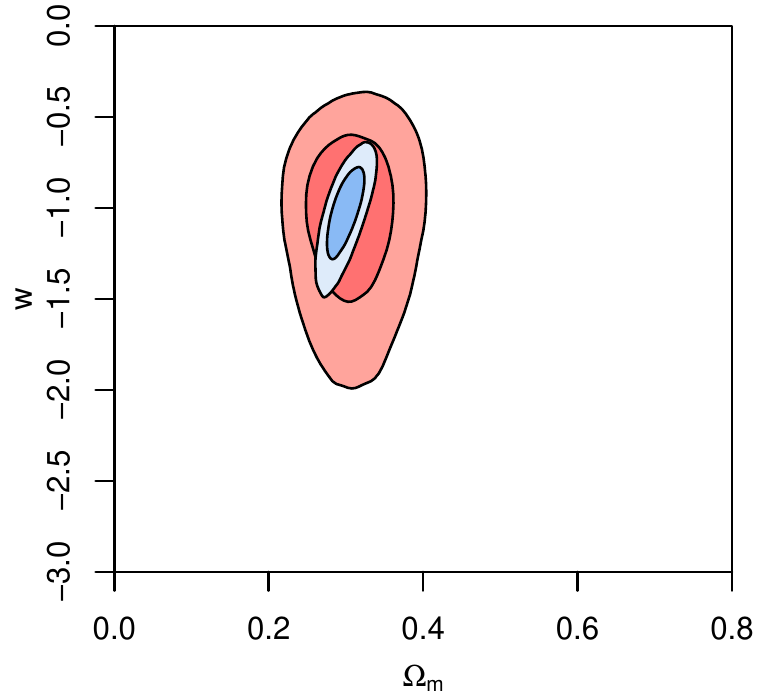}
  \end{minipage}%
  \begin{minipage}[c]{0.33\textwidth}
    \includegraphics[width=\textwidth]{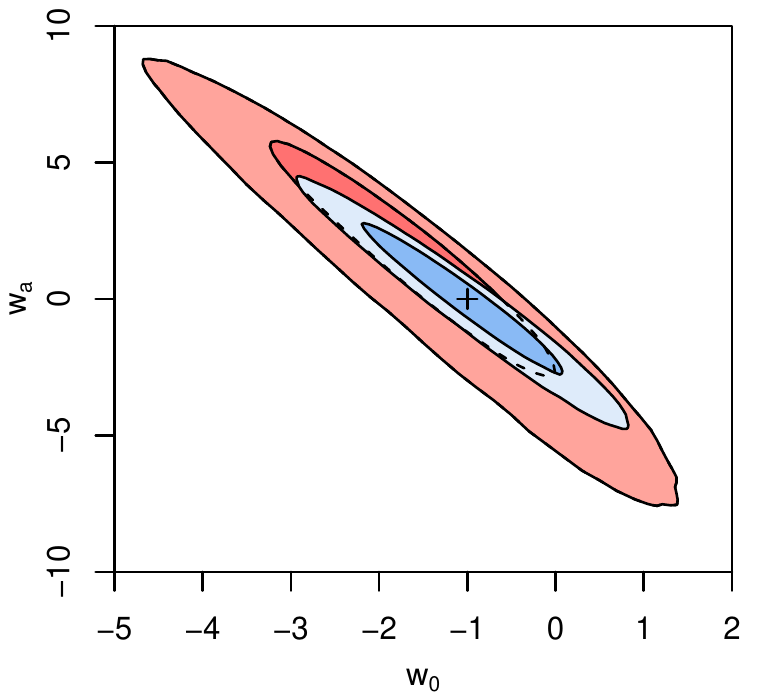}
  \end{minipage}%
  \caption{Joint 68.3\% and 95.4\% confidence constraints from the
\fgas{} test for non-flat \LCDM{} (left), flat $w$CDM (center) and
flat evolving-$w$ (right) models. Red shading shows the constraints
from current \fgas{} data (\citealt{Mantz13}; also employing standard
priors on $\Omegab h^2$ and $h$, as described in the text).  Blue
shading shows the predicted, improved constraints when adding Chandra
data for 53 more clusters, assuming the redshift distribution shown in
Fig.~\ref{fig:zhist}(b) and \fgas{} measurements to $\sim 15$\%
precision.  Optimistic priors are assumed (see Table 1).  The cross in
the right panel marks the cosmological constant model ($w_0=-1$,
$w_a=0$).}
\label{fig:simres}
\end{figure}

\begin{table*}[t]
\begin{center}
\caption{Figures of merit for the \fgas{} experiment, from the combination
of current data and 53 additional clusters observed with Chandra. Our figure of
merit is defined as the inverse of the area enclosed by the 95.4\%
confidence contour for the associated pair of parameters, normalized
by the constraints provided by current data. The last column lists the
fractional constraint on \Omegam{}.}
\label{tab:fom}
\vspace{0.2cm}
\begin{tabular}{ccccc}
  Model & Parameters & Priors & FoM$_{\rm c}$ & $\Delta\Omegam/\Omegam$\\
  \hline                                              
  non-flat \LCDM{} & \Omegam{} -- \Omegal{} & pessimistic & 2.2 &  0.08\\
   &   & optimistic & 5.8 & 0.05\smallskip\\
  flat $w$CDM & \OmegaDE{} -- $w$ & pessimistic & 2.3 & 0.08\\
   &   & optimistic & 6.2 & 0.05\smallskip\\
  flat evolving-$w$ & $w_0$ -- $w_a$ & pessimistic & 2.3 & 0.12\\
  &   & optimistic & 3.5 & 0.09\\
 \hline                      
\end{tabular}
\end{center}
\end{table*}

\noindent Fig.~\ref{fig:simres} shows the predicted constraints from
the simulated Chandra \fgas{} data set (93 clusters; 53 new, 40
existing) for the non-flat \LCDM{}, flat $w$CDM and flat evolving-$w$
models. Future-optimistic priors are assumed. The results from current
\fgas{} data are also shown for comparison. (Note that only cluster \fgas{}
data, in conjunction with priors on $\Omegab h^2$ and $h$, are used
here.) Figures of merit for each model in the future-pessimistic and
future-optimistic cases, normalized to the present results, are shown
in Table~\ref{tab:fom}. Note also the tight constraint on $\Omegam$
achievable with the \fgas{} experiment (at the $\sim 5$\% level,
optimistically), which is largely independent of the cosmological
model assumed.  Intuitively, this constraint originates from the fair
sample argument and the normalization of the $\fgas(z)$ curve, as
described by equation~\ref{eq:fgas3}.

To date, XSZ data have only been able to constrain $h$ with all other
cosmological parameters fixed (e.g. \citealt{Bonamente0512349}). For
this highly restricted model the constraints on $h$ from our simulated
XSZ data set, which involves 500 clusters measured to 20\% precision,
approach the systematic limit imposed by $k(z)$, i.e. $\Delta h \geq
2\Delta k_0$ (considering $k_0$ alone).  Interestingly, the simulated
XSZ data are also able to constrain \Omegam{}, albeit weakly.  The
left panel of Fig.~\ref{fig:xszsim} shows the joint constraints on $h$
and \Omegam{} for a flat \LCDM{} model.  Table~\ref{tab:xsz}
summarizes the constraints on $h$ for this and other models, including
non-flat \LCDM{} and flat $w$CDM.  (These constraints would be
stronger if the intrinsic scatter in XSZ measurements turns out to be
smaller than we have assumed.)

\begin{figure}[t] \centering
  \begin{minipage}[c]{0.4\textwidth}
    \includegraphics[width=\textwidth]{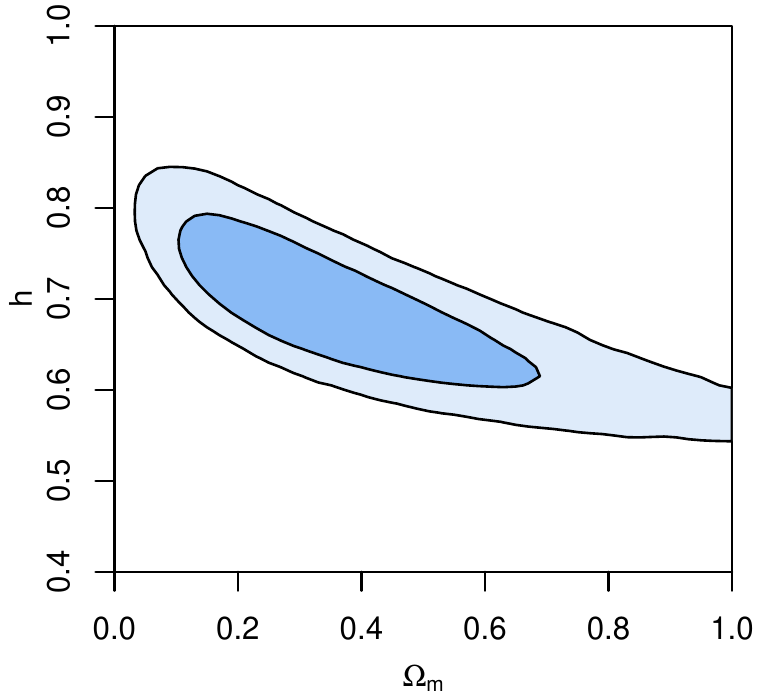}
  \end{minipage}%
\hspace{1.0cm}
  \begin{minipage}[c]{0.4\textwidth}
    \includegraphics[width=\textwidth]{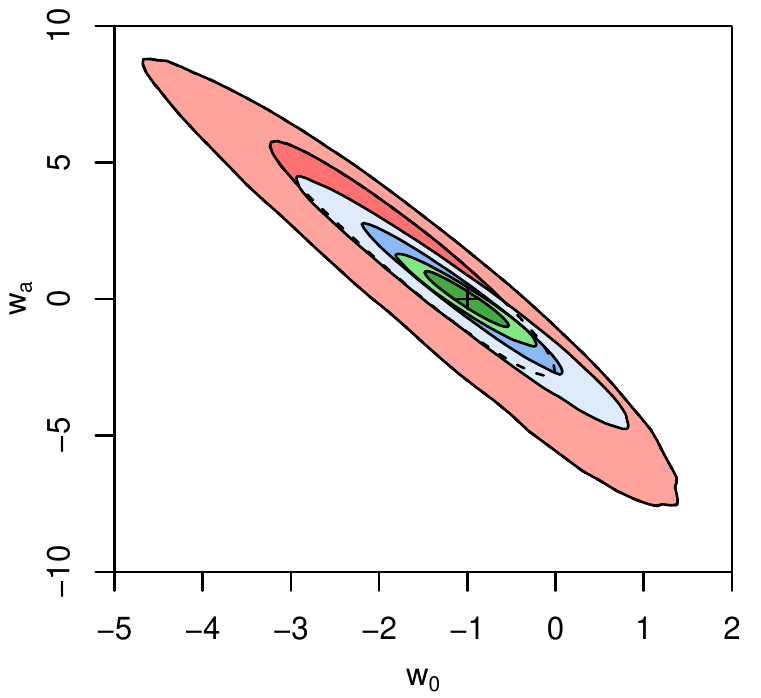}
  \end{minipage}%
  \caption{(\textit{a; left panel}) Predicted 68.3\% and 95.4\%
confidence constraints on the Hubble parameter and mean matter density
from a future XSZ experiment, utilizing relatively short X-ray
observations and follow-up mm-wavelength data for 500 clusters. A flat
\LCDM{} model and optimistic priors (see text) are assumed.
(\textit{b; right panel}) Predicted 68.3\% and 95.4\% confidence
constraints for a more ambitious, future \fgas{} experiment using a
next-generation X-ray observatory with 30 times the collecting area of
Chandra (green inner curves; the outer red and blue curves are
reproduced from Fig.~\ref{fig:simres}). The experiment involves deep
X-ray observations for 450 clusters, providing \fgas{} measurements to
$7.5$\% precision, and assumes 5\% intrinsic scatter.  A flat,
evolving-$w$ model and optimistic priors (see text) are assumed.}
\label{fig:xszsim}
\end{figure}

\begin{table*}[t]
\begin{center}
\caption{Fractional, marginalized constraints on $h$ ($\Delta h/h$) provided by the simulated XSZ data.}
\label{tab:xsz}
\vspace{0.2cm}
 \begin{tabular}{ccc}
   Model & \multicolumn{2}{c}{$\Delta h/h$} \\
   & pessimistic & optimistic \\
   \hline                                              
   only $h$ free & 0.11 & 0.05 \\
   flat \LCDM{} & 0.12 & 0.09 \\
   non-flat \LCDM{} & 0.13 & 0.09 \\
   flat $w$CDM & 0.18 & 0.16 \\
   \hline                      
 \end{tabular}
\end{center}
\end{table*}

\vspace{0.5cm}
\subsection{\fgas{} measurements with a next generation X-ray observatory}
\vspace{0.2cm}

\noindent The program of observations described above, which involves
$\sim 5\%$ of the available Chandra observing time over the next
decade ($\sim 10$\,Ms), is ambitious but achievable given sufficient
community support. However, expanding the scope of this work yet
further will be difficult, with observing time on flagship X-ray
observatories likely to be the limiting factor.  Longer term, the full
exploitation of the \fgas{} technique, which might involve e.g.
doubling the precision of the individual measurements (to approach the
systematic limit) and quadrupling the sample size considered here,
will require a new flagship X-ray observatory with greater collecting
area than Chandra. Possibilities include the SMART-X mission
(http://hea-www.cfa.harvard.edu/SMARTX/) and ATHENA+
\citep{Nandra1306.2307}.

To explore the potential of such an observatory for \fgas{} work, we
have simulated a data set that could be gathered by an observatory
with comparable spatial resolution to Chandra but $\sim 30$ times the
collecting area (other instrument characteristics are assumed to be
the same, in the pessimistic case).  The target clusters are again
drawn fairly from the distribution shown in Fig.~\ref{fig:zhist}a,
with the requirement that $z\geq0.3$. In this case, however, the
simulated \fgas{} measurements have a precision of 7.5\%,
approximately matching the observed level of the intrinsic scatter in
current data.  Adopting a 10\,Ms observing budget, more than 400 new
clusters could be observed, providing a total sample of $\sim 450$ clusters
with \fgas{} measured to this precision.

Our pessimistic projections for this future observatory assume an
intrinsic scatter in \fgas{} of 7.5\%.  In the optimistic case, we
assume that measurements of bulk and turbulent gas velocities with
high resolution X-ray spectrometers will allow us to reduce this
scatter to 5\%. (Bulk and turbulent motions are expected to contribute
significantly to the observed intrinsic scatter in \fgas{}; e.g.
\citealt{Nagai0609247,Battaglia1209.4082}.)

The cosmological constraints for a flat, evolving-$w$ model in the
optimistic case are shown in the right panel of
Fig.~\ref{fig:xszsim}. (Again, only cluster \fgas{} data and priors on
$\Omegab h^2$ and $h$ are used.) Figures of merit, normalized to
the present results, are shown in Table \ref{tab:fom2}, and are a
further factor of 6--7 tighter than those for the 10\,Ms Chandra
program described above.

Given the substantial expected intrinsic scatter in XSZ measurements, the
impact of a new X-ray observatory on that test is likely to be 
relatively modest and we do not consider it further here.

\begin{table*}[t]
\begin{center}
\caption{As for Table~\ref{tab:fom}, but for \fgas{} measurements made
with a future X-ray observatory. The simulated data set includes 450
clusters with \fgas{} measured to 7.5\% precision, and assumes an
intrinsic scatter of $5.0$\%.}
\label{tab:fom2}
\vspace{0.2cm}
\begin{tabular}{ccccc}
  Model & Parameters & Priors & FoM$_{\rm c}$ & $\Delta\Omegam/\Omegam$\\
  \hline                                              
  flat evolving-$w$ & $w_0$ -- $w_a$ & pessimistic & 16.7 & 0.10\\
  &   & optimistic & 22.6 & 0.06\\
 \hline
\end{tabular}
\end{center}
\end{table*}

\vspace{0.5cm}
\section{Conclusions}
\vspace{0.2cm}

\noindent We have examined the ability of current and near-term
multiwavelength observations of galaxy clusters to measure cosmic
distances and constrain cosmology via the \fgas{} and XSZ
experiments. Existing \fgas{} measurements for hot ($kT>5$\,keV),
dynamically relaxed clusters ($\sim 40$ systems, mostly at $z\leq0.6$;
\citealt{Mantz13} and references therein) provide competitive
constraints on dark energy, comparable to those from current BAO and
SN\,Ia studies. Expanding this work to include measurements for
approximately 100 such clusters, spanning the redshift range
$0<z<1.4$, with \fgas{} measured to a precision of 15\% in the
0.8--1.2\,$r_{2500}$ shell, should constrain dark energy with a figure
of merit $4-6$ times better, depending on the cosmological model. In
particular, the \fgas{} data provide a tight constraint on $\Omegam$,
independent of the cosmological model assumed. Our projections assume
modest improvements in hydrodynamical simulations and weak lensing
mass calibration, and the availability of approximately 5\% of the
total Chandra observing budget ($\sim 10$\,Ms) for this work over the
next decade, noting that these Chandra observations will also
facilitate a broad range of astrophysical studies. We assume a
redshift distribution of the observed clusters that is representative
of the population that eROSITA will find, although in practice this
redshift distribution could be further optimized for dark energy
studies.

In determining our predicted dark energy constraints we have employed
the same MCMC method used to analyze current data.  This encapsulates
all of the relevant degeneracies between parameters and allows one to
easily and efficiently incorporate priors and systematic allowances in
the analysis.

The availability of observing time on major X-ray observatories such
as Chandra and XMM-Newton is likely to be the limiting factor for the
$\fgas(z)$ experiment.  Longer term, the full exploitation of this
technique will require a new flagship X-ray observatory with
comparable spatial resolution but greater collecting area than
Chandra.  We have considered the prospects for distance measurements
with such an observatory, demonstrating that the constraints on
evolving dark energy models are likely to be improved by a further
factor of $6-7$.

We conclude that the \fgas{} and XSZ experiments offer a powerful
approach for measuring cosmic distances which should be competitive
with and complementary to other leading methods over the next decade.

\vspace{0.5cm}
\section*{Acknowledgments}
\vspace{0.1cm}

\noindent This work was supported in part by the US Department of
Energy under contract number DE-AC02-76SF00515, by the 
National Aeronautics and Space Administration through 
Chandra Award Numbers GO8-9118X and TM1-12010X, and by the 
National Science Foundation under grants AST-0838187 and AST-1140019.
The Dark Cosmology Centre (DARK) is funded by the Danish National 
Research Foundation. 

\vspace{0.5cm}

\bibliography{references}

\end{document}